# Quantum Interference and Optical Tuning of Self-Trapped Exciton State in Double Halide Perovskite

Kai-Xuan Xu[1,2,#], Xin-bao Liu[3,7,#], Simin Pang[1,2,#], Zhe Zhang[1,2], Yubin Wang[5], Jiajun Luo[6], Jiang Tang[6], Qihua Xiong[5], Sheng Meng[3,7,*], Shiwu Gao[4,†], and Jun Zhang[1,2,‡]

[1]*State Key Laboratory of Superlattices and Microstructures, Institute of Semiconductors, Chinese Academy of Sciences, Beijing 100083, China*

[2]*Center of Materials Science and Optoelectronics Engineering, University of Chinese Academy of Sciences, Beijing 100049, China*

[3]*Beijing National Laboratory for Condensed Matter Physics and Institute of Physics, Chinese Academy of Sciences, Beijing 100190, China*

[4]*Beijing Computational Science Research Center, Beijing 100193, China*

[5]*State Key Laboratory of Low-Dimensional Quantum Physics and Department of Physics, Tsinghua University, Beijing, 100084, China*

[6]*Wuhan National Laboratory for Optoelectronics (WNLO), School of Optical and Electronic Information, Huazhong University of Science and Technology (HUST), Wuhan 430074, China*

[7]*School of Physical Sciences, University of Chinese Academy of Sciences, Beijing 100190, China*

[#]These authors contributed equally.

**Corresponding Author**

*E-mail: [‡]zhangjwill@semi.ac.cn; [†]swgao@csrc.ac.cn; [*]smeng@iphy.ac.cn

**ABSTRACT.** Self-trapped excitons (STEs), renowned for their unique radiative properties, have been harnessed in diverse photonic devices. Yet, a full comprehension and manipulation of STEs remain elusive. In this study, we present novel experimental and theoretical evidence of the hybrid nature and optical tuning of the STEs state in $Cs_2Ag_{0.4}Na_{0.6}InCl_6$. The detection of Fano resonance in the laser energy-dependent Raman and photoluminescence spectra indicates the emergence of an exciton-phonon hybrid state, a result of the robust quantum interference between the discrete phonon and continuous exciton states. Moreover, we showcase the ability to continuously adjust this hybrid state with the energy and intensity of the laser field. These significant findings lay the foundation for a comprehensive understanding of the nature of STE and its potential for state control.

## I. INTRODUCTION.

Electron-phonon coupling significantly affects materials' electrical conductivity [1,2], superconductivity [3,4], and optical absorption properties [2,5]. When this coupling is sufficiently strong, an electron can distort the surrounding lattice, creating a localized state known as a polaron [6-8]. In this case, the electron moves with a local lattice distortion, resembling being "dressed" by phonons, which affects its effective mass and mobility [9-12]. Polarons play a crucial role in understanding charge transport [13] and optical emission within specific materials [14], such as semiconductors [15], oxides [7], and certain ionic crystals [16]. A quasi-particle excitation known as self-trapped exciton (STE) forms when a polaron traps an electron-hole pair (exciton) in a potential well created by lattice distortions, resulting in a stable state that influences radiative recombination [17-19]. Due to their unique radiative properties, STEs have been harnessed in photonic devices like lasers [20] and white-light emitters [17]. Therefore, comprehending the mechanisms facilitating self-trapping and manipulating these states for targeted applications becomes essential.

The essence of exciton-phonon coupling in STEs lies in a quasiparticle composed of an exciton cloaked by a cloud of vibronic phonons, forming a hybrid quantum state that is part phonon, part exciton. Although optical spectroscopies have shed light on the temporal and spatial scales of the photoexcitation-induced reorganization of electronic bands [21,22], and lattice vibrations [23-25], the precise nature of the hybrid STE state remains elusive. In principle, the nature of the hybrid STE state can be directly proved by resonant Raman and resonant photoluminescence (PL) techniques, where the phonon frequency and STE energy should exhibit resonance quantum interference. Resonance quantum interference is a general phenomenon, which strongly affects the electronic transport, optical, and vibronic properties of materials [2,26,27]. As one of the most representative phenomena of quantum interference, Fano resonance describes interference between continuum states and discrete states, making it an ideal platform for studying the strongly interacting physics [2,26,28], such as the magnetization and electronic polarization [29-31], resonant electromagnetic effects [32], and exciton-phonon interactions (EPIs) [33-35]. In particular, Fano resonance in Raman scattering induced by EPIs provides a powerful tool to reveal underlying physics in solid materials [2,34,35]. Among the most promising STEs materials, lead-free double perovskites [36] stand out for their environmental friendliness and high stability, showcasing considerable potential in various applications, including white light sources [37], photodetectors [38], solar cells [39], and X-ray detectors [40]. In this study, we present theoretical and experimental evidence elucidating the nature and formation dynamics of an exciton-phonon hybrid STE state in the halide double perovskite $Cs_2Ag_{0.4}Na_{0.6}InCl_6$ and demonstrate the effective modulation of the STE state using a steady-state optical field. Remarkably, the STE emission energy (linewidth) and the associated vibrational frequency (linewidth) display a Fano resonance as a function of excitation energy, indicating the many-body quantum interference characteristic of the exciton-phonon hybrid state. Furthermore, we find that the excitation power can modify the excited state potential and EPI strength, thus facilitating the optical tuning of STE emission and phonon frequency. Our findings reveal the intrinsic nature of the exciton-phonon hybrid state in STEs and establish optical tuning via a steady-state optical field.

## II. METHODS

### A. Raman, PL, and PLE spectra measurements

The $Cs_2Ag_{0.4}Na_{0.6}InCl_6$ crystal with 0.04% bismuth doping was synthesized using a hydrothermal method. More details about the sample preparation can be found elsewhere [17]. The Raman and a contour plot of PLE spectra are obtained from Jobin-Yvon HR800, T64000 systems and Edinburgh FLS-1000 fluorescence spectrophotometer, respectively, the experimental details for obtaining spectra can be found in Supplemental Material [41].The Raman spectra resolutions are also available in Supplemental Material [41].

### B. First-principles calculations

First-principles calculations were performed using VASP with the PAW method and PBE0 hybrid functional to mitigate self-interaction errors [42-44]. The phonon structure was obtained using the PBEsol functional [45,46], yielding accurate agreement with experimental data. Time-dependent density functional theory (TDDFT) combined with Ehrenfest molecular dynamics was used to compute the dynamics of the $A_{1g}$ mode and STEs, with further details available in prior work [47]. More details on the computational methods can be found in Supplementary Material [41].

## III. RESULTS AND DISCUSSION

The formation dynamics of STE and its dynamic phonon evolution are schematically described in Fig. 1. Upon photoexcitation, an electronic transition occurs from the ground state to the free exciton state, leading to the charge transfer from Ag to In atoms within the lattice. As shown in the upper panel of Fig. 1(a), the charge density difference between the electronic excited state and ground state reveals that the excited holes are preferentially located in the Ag atoms. Due to the strong EPIs, the excited electrons generate coherent phonon oscillations, characterized by the $A_{1g}$ mode during the first few hundred femtoseconds [48]. The polarization of the $A_{1g}$ mode is schematically shown in the right panel of Fig. 1(b). The lattice distortion mainly manifests itself in the change of bond length, where the bond lengths between atoms in the *a*-*b* plane of the lattice are squeezed, and those along the *c* axis are simultaneously stretched [49]. Subsequently, the excited electrons were trapped by the corresponding lattice deformation potential through the excited-state structure reorganization, leading to the STE state's formation. At this time, the electron-hole pairs are localized after the trapping process mediated by electron-phonon coupling [50], where the holes are pinned on the Ag atom sites and screened by the surrounding electrons [lower panel of Fig. 1(a)]. Finally, the radiative recombination of the STE generates a broad emission spectrum [Fig. 2(a)], and then the whole system returns to its ground state [51]. To further understand the transient changes in the $A_{1g}$ mode, Fig. 1(c) shows the time-resolved atomic displacements ($Q$) of the $A_{1g}$ mode calculated by TDDFT combined with the Ehrenfest molecular dynamics simulation before and after photoexcitation. Before photoexcitation, the $A_{1g}$ mode vibrates with a small amplitude around the equilibrium position. After photoexcitation [indicated by an orange shadow in Fig. 1(c)], a larger amplitude of oscillation is generated and shifted to a new equilibrium position (indicated by the orange dashed line) within a few hundred femtoseconds [52]. This shifted lattice structure exactly agrees with the lattice distortion induced by the localized electron-hole pair as shown in the lower panel in Fig. 1(a), indicating the occurrence of lattice distortion induced by trapped extra electrons. In particular, with the resonant excitation, due to the Fröhlich interaction [25,49], the STE strongly couples with the phonon around the lattice site, forming the exciton-phonon hybrid state. The STE and corresponding coherent phonon formation process suggest that altering the wavelength and power of excitation lights is a potential method to tune the exciton-phonon hybrid state.

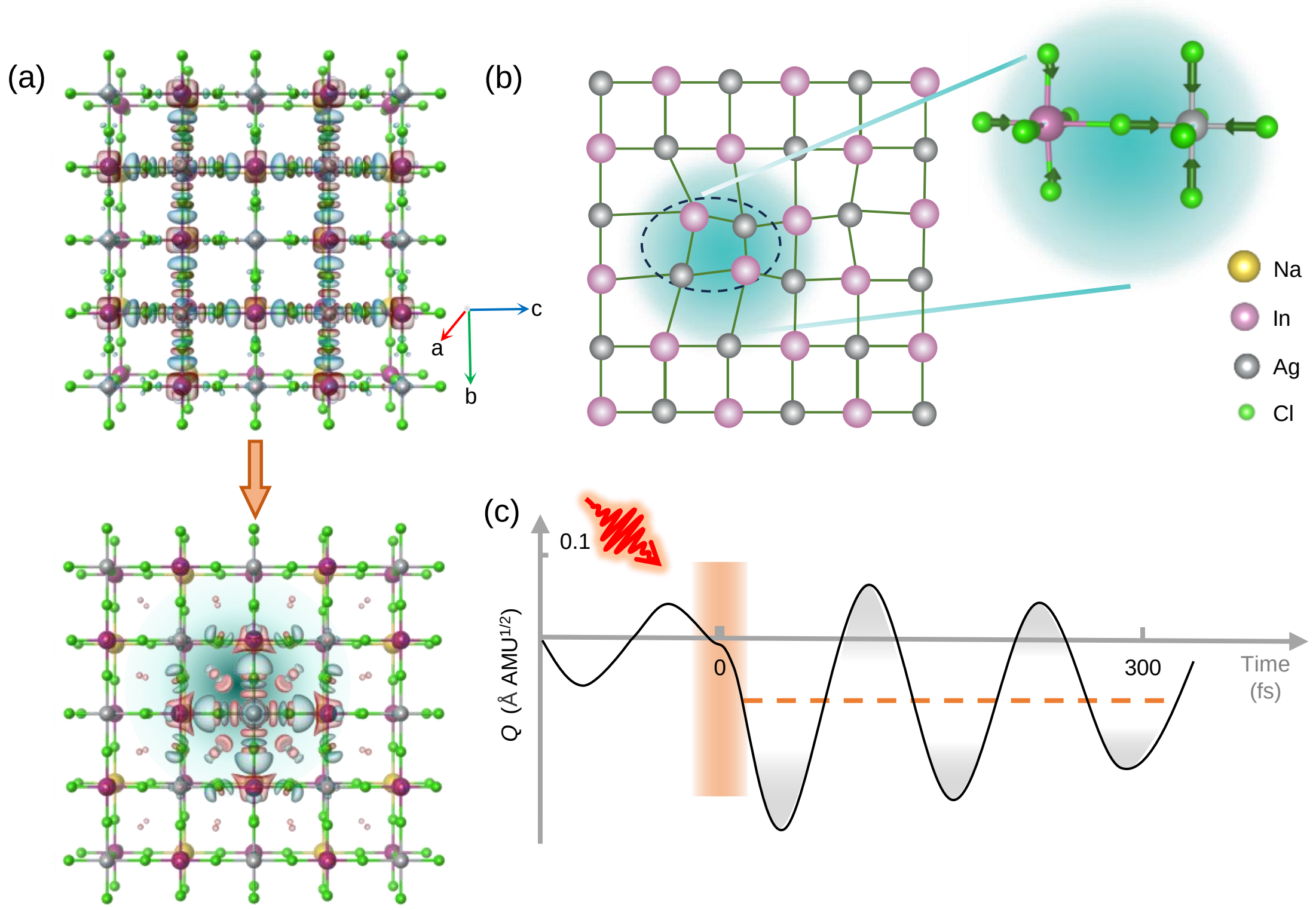

FIG. 1. The STE formation and corresponding lattice distortion of $Cs_2Ag_{0.4}Na_{0.6}InCl_6$. (a) The differential charge density distribution between photoexcited and ground state before (upper panel) and after (lower panel) the charge trapping process. The red and cyan isosurfaces represent the electron and hole orbital densities, respectively. The upper panel shows the charge transfer from Ag and Cl atoms to In atom after photoexcitation. The lower panel shows that the electron-hole pairs are localized after the trapping process mediated by electron-phonon coupling, where the holes are pined on the Ag sites and screened by spreading electrons. (b) The lattice distortion and the atomic vibration distortion schematic diagram when forming the STE state. The Cl atoms are omitted in the diagram for clarity. (c) The time-resolved atomic displacement ($Q$) of the $A_{1g}$ phonon mode before and after the photoexcitation (black line). Before the laser pulse illumination, $Q$ shows a small amplitude oscillation around the origin equilibrium position and then has a larger amplitude after photoexcitation around a new equilibrium position (orange dashed line).

The $Cs_2Ag_{0.4}Na_{0.6}InCl_6$, whose average emission energy $E_{PL}$ (2.18 eV, 570 nm) is derived from the energy difference between the ground state and the excited state after forming the STE state, is a direct bandgap semiconductor[41], in agreement with the reported result [53]. As shown in Fig. 2(a), the photoluminescence excitation (PLE) spectrum of $Cs_2Ag_{0.4}Na_{0.6}InCl_6$ obtained for the 608 nm emission indicates that the resonant excitation occurs at 355 nm. In contrast, 532 nm excitation cannot generate the STE state. The broad emission spectrum with a mean emission wavelength of about 608 nm can be well recognized from the PL spectrum under resonant 355 nm excitation. Besides the exciton emission, the characteristics of phonon modes also depend on the formation of the STE state. For comparison, we first investigate the phonon modes without forming STE by performing polarized Raman spectroscopy measurements using non-resonant 532 nm excitation. As shown in Fig. 2(b), two kinds of Raman-active phonon modes, namely the $A_{1g}$ and $E_g$ modes, are identified. The schematic diagrams of the phonon eigenvectors are depicted in the insets of Fig. 2(b). Considering the random distribution of $Na^+$ ($Ag^+$) in the lattice sites, $Cs_2Ag_{0.4}Na_{0.6}InCl_6$ can be attributed to the $O_h$ point group. By combining phonon dispersion calculated by density functional perturbation theory (DFPT) [41], Raman tensors [41], and the polarized Raman spectra, the peaks at around 50 $cm^{-1}$ and 141 $cm^{-1}$ are identified as $E_g$ vibration modes of the $(BCl_6)^{5-}$ (B= Ag, In, and Na) octahedrons, with the Cl atoms around $B^+$ atoms vibrating in a different direction. The peak at around 296 $cm^{-1}$ is assigned to the $A_{1g}$ vibration mode of the $(BCl_6)^{5-}$ octahedrons, where the Cl atoms around $B^+$ atoms simultaneously vibrate in phase. The larger ionic radius of $Ag^+$ compared to $In^+$ results in a lattice mismatch, making the Ag-Cl bond stiffer than the In-Cl bond. This suggests that the $A_{1g}$ and $E_g$ phonon modes primarily originate from the octahedron vibrations of $(AgCl_6)^{5-}$ in the perovskite structure. Based on the Raman tensors, both the $A_{1g}$ and $E_g$ modes would disappear under the cross (HV) polarization configuration. However, we can still observe the weak $A_{1g}$ and $E_g$ modes under the HV polarization configuration. A possible reason is that the slight lattice distortion induced by the Na cation doping would lead to the deviation of the Raman tensors, relaxing the Raman selection rules [54].

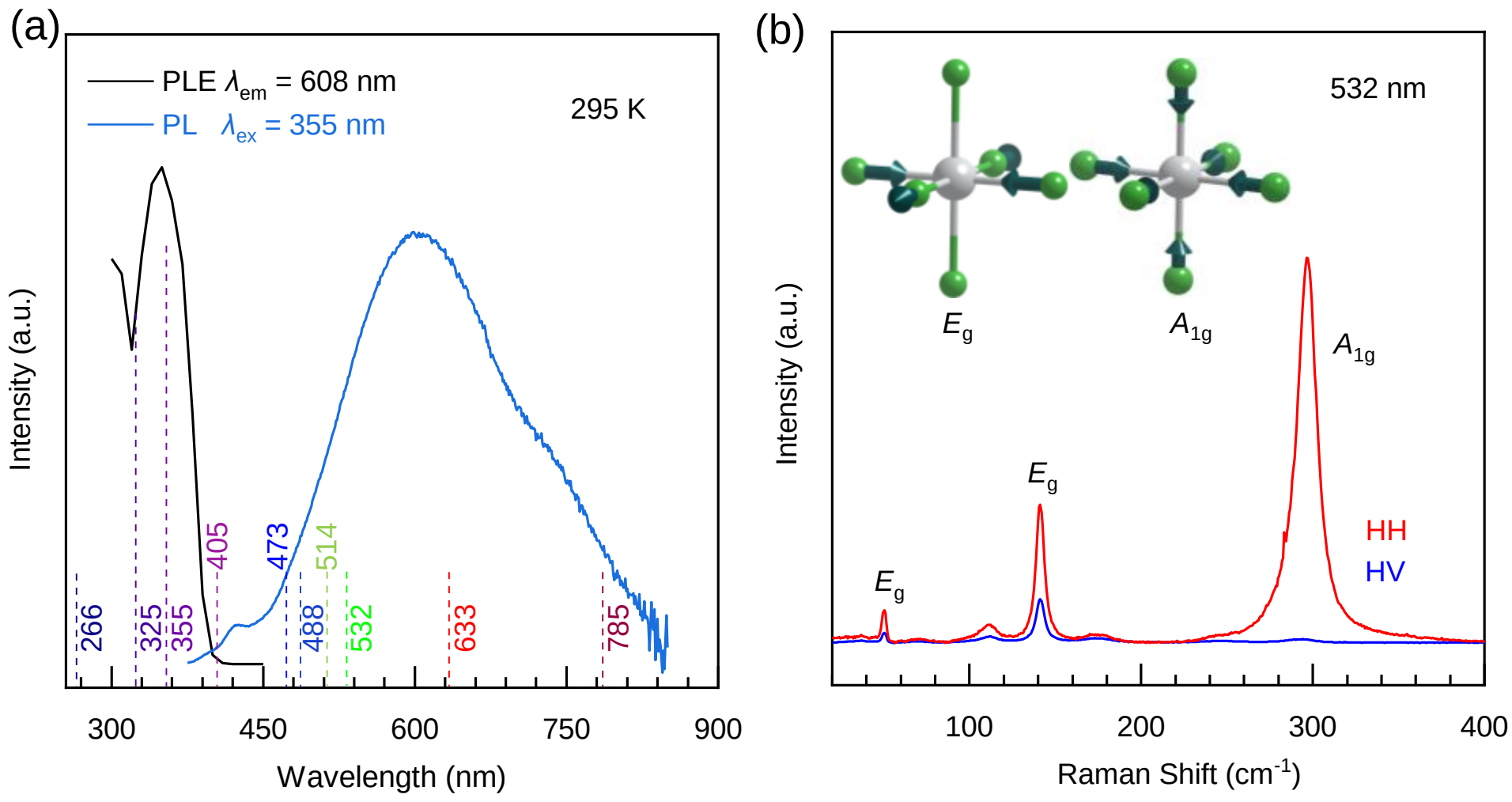

FIG. 2. The exciton emission and lattice vibration of $Cs_2Ag_{0.4}Na_{0.6}InCl_6$. (a) The PL and PLE spectra of $Cs_2Ag_{0.4}Na_{0.6}InCl_6$. As mentioned in the main text, the dotted lines indicate different excitation wavelengths used in the Raman measurements. (b) The polarized Raman spectra by 532 nm excitation, with the insert showing the vibration of the $E_g$ and $A_{1g}$ modes, respectively. HV and HH represent the cross-polarization and parallel-polarization configurations of incident and scattered light, respectively.

To investigate how the STE formation affects the phonon mode, we performed Raman spectroscopy measurements using excitations both below and above the resonant energy level. Figure 3(a) shows the Raman spectra obtained with ten different excitation energies (wavelengths) ranging from 1.58 eV (785 nm) to 4.66 eV (266 nm), revealing a noticeable redshift of the $A_{1g}$ mode under 355 nm resonant excitation. STE formation during resonant excitation induces a strong charge transfer, which changes the Ag-Cl bond and exciton parity, leading to significant lattice distortion and subsequent modifications of the phonon modes. Besides, the $E_g$ mode at around 141 $cm^{-1}$ disappears completely under 325 and 355 nm excitations, indicating its instability during the STE formation. Due to this instability of the $E_g$ peak at resonant excitation, it is difficult to track its dynamic behavior. Therefore, we focused on the more stable $A_{1g}$ mode and investigated its tuning mechanism.

Owing to the "soft" lattice and strong EPI [6,55], phonon and exciton are not independent in the system but form an exciton-phonon state with hybrid characteristics. Such a hybrid state strongly modifies the frequency and damping of the phonons around the exciton resonant energy level [28,34]. As displayed in Fig. 3(a), the $A_{1g}$ modes show an obvious frequency softening and linewidth broadening around the resonant excitation. The Raman shift and full width at half maximum (FWHM) extracted from Fig. 3(a) can be well fitted by a Fano function [red and blue curves in Fig. 3(b)], which reveals a many-body quantum interference behavior involving photon, exciton, and phonon. Typically, the Fano lineshape arises from the quantum interference between the discrete state and continuum state [2,26,28,34,35]. Since the linewidth of exciton is much larger than that of phonon, the discrete state is the $A_{1g}$ phonon mode, and the continuum state is the exciton state. The Fano lineshape signifies that the excitation energy can tune the exciton-phonon hybrid state. Under the Fano resonance condition, the phonon frequency $\omega(E_l)$ at the excitation energy $E_l$ has a lineshape described by [28,34,35]:

$$\omega(E_l) - \omega_0 = \frac{\Delta\omega\left(1+\frac{E_l-E_{ex}}{q\Gamma}\right)^2}{1+\left(\frac{E_l-E_{ex}}{\Gamma}\right)^2}, \qquad (1)$$

where $\omega_0$ is the frequency of the bare $A_{1g}$ mode, $E_{\rm ex}$ and $\Gamma$ is the energy and damping of the bare free exciton, respectively. $E_l$ is the excitation energy, $\Delta\omega$ is the softening ($\Delta\omega < 0$) or stiffening ($\Delta\omega > 0$) value of phonon frequency at resonance, and the dimensionless parameter $q$ characterizes the relative dipole strength of the renormalized phonon and exciton. Depending on the value and sign of $q$, the Fano lineshape could describe resonance ($|q| \gg 1$, exciton dominates), anti-resonance ($|q| \ll 1$, phonon dominates), and dispersion ($|q| \approx 1$, comparable phonon and exciton contribution) situations [34,35]. Here, the fitting parameters for the frequency of the $A_{1g}$ mode are $\omega_0 = 298$ cm$^{-1}$, $\Delta\omega = -8$ cm$^{-1}$, $E_{\rm ex} = 3.44$ eV, $\Gamma = 0.11$ eV, and $q = 2.5$. The value of $q$ indicates that comparable phonon and exciton contributions exist in the hybrid state, which leads to the renormalization of the frequency and linewidth of the $A_{1g}$ mode. As the excitation energy approaches the resonant energy level (355 nm), a pronounced dip is observed, indicating the greatest extent of frequency softening relative to other excitation wavelengths. Due to exciton screening in non-resonant excitation, the $A_{1g}$ mode frequency is slightly lower than the bare phonon frequency in the ground state. Similarly, the FWHM of the $A_{1g}$ mode as a function of excitation energy also shows a Fano profile with similar fitting values of $E_{\rm ex}$, $\Gamma$, and $q$ mentioned above [blue curve in Fig. 3(b)], reaching a maximum under 355 nm excitation. In the absence of STE formation, the FWHM of the bare $A_{1g}$ mode is about 12 cm$^{-1}$, while resonant excitation results in an increase in FWHM of about 10 cm$^{-1}$.

Since the hybrid state shows the half-exciton-half-phonon characteristic, tuning of the hybrid state involves not only the modulation of the phonon mode as mentioned above, but also the manipulation of the exciton emission, which is manifested in the change of the PL linewidth and central wavelength. As shown in Fig. 3(c), the FWHM and mean wavelength of the PL spectra for STE (extracted from Supplemental Material [41] [56]) versus excitation energy also exhibit the characteristics of Fano lineshape. Here, the fitting parameters are $\lambda_0 = 622$ nm, $\Delta\lambda = -17.5$ nm, $E_{\rm ex} = 3.65$ eV, $\Gamma = 0.28$ eV, and $q = 4.3$ for the central wavelength, while $\mathrm{FWHM}_0 = 612$ meV, $\Delta\mathrm{FWHM} = 42$ meV, $E_{\rm ex} = 3.65$ eV, $\Gamma = 0.28$ eV, and $q = 3$ for the linewidth. The $q$ here agrees with the fitting results obtained from the Raman spectra, further indicating the existence of the exciton-phonon hybrid state with comparable contributions from phonon and exciton. Our experimental results demonstrate that optical excitation can efficiently tune the STE emission and phonon modes in the hybrid state.

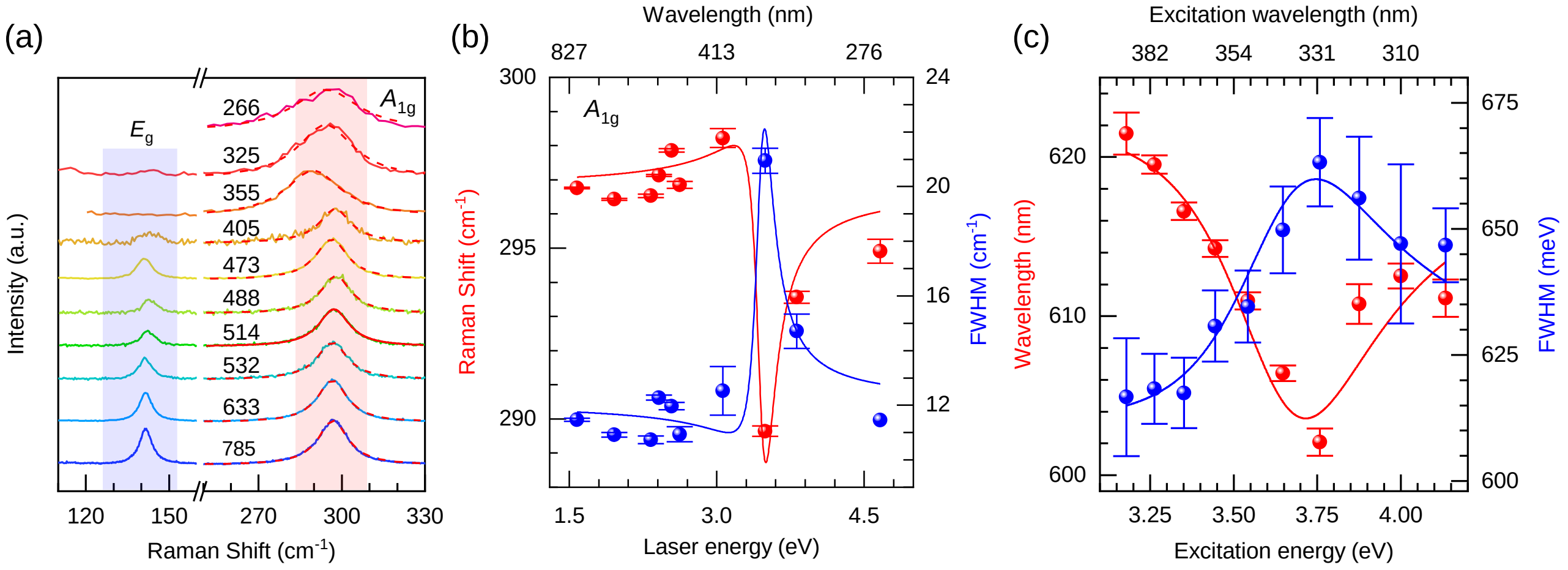

FIG. 3. The renormalization effects of the $A_{1g}$ mode and the STE emission for $Cs_2Ag_{0.4}Na_{0.6}InCl_6$. (a) Normalized Raman spectra with ten excitation wavelengths. (b) The Raman shifts and FWHMs of the $A_{1g}$ mode as a function of the excitation energy, extracted from (a). (c) The center wavelengths, and FWHMs of the PL spectra as a function of the excitation energy. The red and blue curves in b and c are fitting results using the Fano function, and the error bars correspond to the fitting error.

For the exciton-phonon hybrid state, alternating the exciton concentration is an effective method to adjust the EPI and then the lattice distortion, further causing the phonon self-energy renormalization [21] and modifying the phonon frequency and damping [22]. As shown in Supplemental Material [41], to evaluate the EPI strength, we theoretically calculated the excited state potential and Huang-Rhys factor ($S$ factor) with increasing carrier concentration $n$, respectively [57]. The $S$ factor exhibits an increase proportional to $n^2$, indicating the enhanced EPI with higher carrier concentration. To investigate how the carrier concentration further alters the lattice distortion, we calculated the variations of the bond lengths for Ag-$Cl_{x,y}$, Ag-$Cl_z$, Na-$Cl_{x,y}$, and Na-$Cl_z$ with increasing carrier concentration. The variations of the bond lengths are defined as the difference between the bond lengths with and without electron doping. As shown in Fig. 4(a), we observed apparent variations of the bond lengths along with the Ag-Cl bonds, and only the Ag-$Cl_z$ bond is stretched with the increasing carrier population. Since the dominant lattice distortion direction is consistent with the vibrational direction of the Raman-active $A_{1g}$ mode, the apparent softening of the $A_{1g}$ mode can be observed when forming the STE state. To intuitively understand the frequency softening of the $A_{1g}$ mode, we theoretically calculated the frequency of the $A_{1g}$ mode as a function of the carrier concentration $n$, as shown in Fig. 4(b). The calculated frequency gradually reduces with increasing $n$, satisfying $\omega_c = 292.76 - 2.29\sqrt{n}$. We performed the power-dependent Raman spectra measurements under the resonant 355 nm excitation to confirm the theoretical result. As shown in Fig. 4(c), the pronounced decrease in frequency and increase in FWHM of the $A_{1g}$ mode can be observed with increasing excitation light power. The Raman shift and FWHM can be fitted using $\omega = 290.36 - 2.73\sqrt{P}$ and $FWHM = 20.71 + 6.33\sqrt{P}$, respectively. On the contrary, under the 532 nm excitation, both the Raman shift and FWHM remain almost unchanged with the increasing power [41]. The full power-dependent spectra under 532 nm and 355 nm excitations are shown in Supplemental Material [41]. Since the excitation power ($P$) is directly proportional to the carrier concentration ($n$) within the power range used in the experiments, the theoretical result for the $A_{1g}$ frequency agrees well with the experimental one, which exhibits the $\sqrt{n}$ (or $\sqrt{P}$) dependence. The $\sqrt{n}$ dependence of phonon frequency is similar to the case of plasma frequency (proportional to $\sqrt{n_e}$) with screening effect. For small $q$, in Thomas-Fermi approximation, the oscillation frequency can be described by the following formula [58]:

$$\omega(q)=\sqrt{\frac{4\pi Z e^2 n_e}{M k_{TF}^2}}q, \tag{2}$$

where $Z$ is the number of electrons per primitive unit cell, $n_e$ the electron density, $M$ the ion mass, $k_{TF}^2$ the Thomas-Fermi wave number, and $q$ is the wavevector. The similar $\sqrt{n}$ dependence indicates that, like the screening effects electrons have on the ions, the phonons "dressed" with the exciton cloud also experience the excitonic shielding, further demonstrating the comparable contributions in the exciton-phonon hybrid state. Our results show that the all-optical method, including using excitations with different energy and power, can change the carrier concentration and the EPI parameter [59], further effectively manipulating the hybrid state.

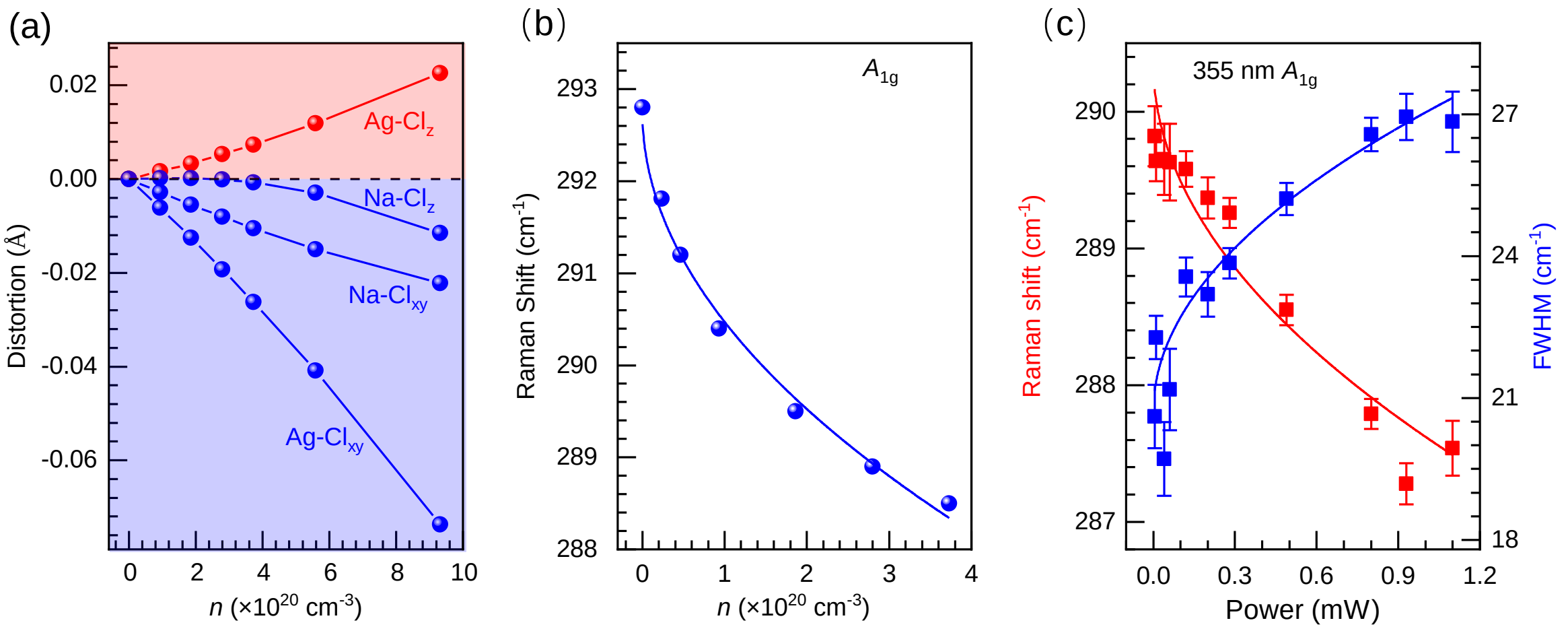

FIG. 4. The calculated variation of bond length and frequency of the $A_{1g}$ mode, together with the excitation power-dependent frequency and FWHM of the $A_{1g}$ mode. (a) The calculated distortion of the bond length for Ag-$Cl_{x,y}$, Ag-$Cl_z$, Na-$Cl_{x,y}$, and Na-$Cl_z$ as a function of the carrier concentration *n*. (b) The calculated frequency of the $A_{1g}$ mode as a function of the carrier concentration *n*. (c) The values of Raman shifts and FWHMs of the $A_{1g}$ mode with increasing excitation power at 355 nm excitation. The error bars correspond to the fitting error. The blue and red solid lines in b and c represent the fitting results.

## V. CONCLUSIONS

Using the resonant Raman and PL spectroscopy techniques, we experimentally demonstrate the existence and manipulation of the exciton-phonon hybrid state in the halide double perovskite $Cs_2Ag_{0.4}Na_{0.6}InCl_6$. The hybrid nature of this state is revealed by observing the Fano lineshape in the variation of phonon and STE emission versus excitation energy. Combining DFT calculations, we identify that the $A_{1g}$ phonon mode dominates the STE formation, and the phonon softening is associated with the increasing carrier concentration and the enhanced EPI. The quantum interference strength between excitons and phonons is strongly determined by the vibration way of phonons, the excitation energy, and the excitation power, thus allowing the tuning of the hybrid state. This tunable coupling between STE, phonon vibrations, and external light fields presents a promising platform for investigating EPI in strongly coupled systems [60,61]. Additionally, the steady-state optical method employed in this study to investigate multi-body quantum interferences can be extended to other systems, and our findings would facilitate the development of optically modulated STE-based devices.

## ACKNOWLEDGMENTS

J. Z. acknowledges the funding support from the Chinese Academy of Sciences - the Scientific and Technological Research Council of TÜRKİYE Joint Research Projects (172111KYSB20210004), the CAS Interdisciplinary Innovation Team, National Natural Science Foundation of China (12074371), and Research Equipment Development Project of Chinese Academy of Sciences (YJKYYQ20210001). The authors thank Prof. Minghui Fan at USTC for her valuable support in conducting the Raman measurements as part of the Grand Technical Talent Promotion Plan (TS2021002).

# Supplementary Material for

# "Quantum Interference and Optical Tuning of Self-Trapped Exciton State in Double Halide Perovskite"

Kai-Xuan Xu[1,2,#], Xin-bao Liu[3,7,#], Simin Pang[1,2,#], Zhe Zhang[1,2], Yubin Wang[5], Jiajun Luo[6], Jiang Tang[6], Qihua Xiong[5], Sheng Meng[3,7,*], Shiwu Gao[4,†], and Jun Zhang[1,2,‡]

[1]*State Key Laboratory of Superlattices and Microstructures, Institute of Semiconductors, Chinese Academy of Sciences, Beijing 100083, China*

[2]*Center of Materials Science and Optoelectronics Engineering, University of Chinese Academy of Sciences, Beijing 100049, China*

[3]*Beijing National Laboratory for Condensed Matter Physics and Institute of Physics, Chinese Academy of Sciences, Beijing 100190, China*

[4]*Beijing Computational Science Research Center, Beijing 100193, China*

[5]*State Key Laboratory of Low-Dimensional Quantum Physics and Department of Physics, Tsinghua University, Beijing, 100084, China*

[6]*Wuhan National Laboratory for Optoelectronics (WNLO), School of Optical and Electronic Information, Huazhong University of Science and Technology (HUST), Wuhan 430074, China*

[7]*School of Physical Sciences, University of Chinese Academy of Sciences, Beijing 100190, China*

[#]These authors contributed equally.

**Corresponding Author**

*E-mail: [‡]zhangjwill@semi.ac.cn; [†]swgao@csrc.ac.cn; [*]smeng@iphy.ac.cn

## 1. The calculated electron band structure and potential energy surface (PES) of $Cs_2Ag_{0.5}Na_{0.5}InCl_6$

All the first-principles density functional perturbation theory (DFPT) calculations were performed using the projector augmented wave method as implemented in the Vienna Ab initio Simulation Package with the PAW potentials [1]. We adopted the PBE0 hybrid functional which can correct self-interaction errors in generalized gradient approximation (GGA) to describe the exchange-correlation interactions [2]. To simulate the effect of Na atom doping, we chose the half-substituted structure $Cs_2Ag_{0.5}Na_{0.5}InCl_6$, as shown in Fig. 1(b), in which Ag atoms alternate with Na atoms to form the layer structure. According to our tests, the fully relaxed structure using the Perdew-Burke-Ernzerhof (PBE) functional underestimates the frequency of $A_{1g}$ mode by about 10.4%. So the phonon structures were calculated using PBEsol functional, which gets excellent agreement with the experimental results [3]. We obtained the lattice parameters $a = b = 10.589$ Å, and $c = 10.662$ Å. Owing to this large supercell, a $3 \times 3 \times 3$ k-mesh is sufficient to sample the Brillouin zone. The plane-wave energy cutoff was set to 800 eV, and the atomic forces were converged to within $10^{-5}$ Ry/a.u. Constrained density functional theory (DFT) was used to calculate the softening of $A_{1g}$ mode and the potential energy surface (PES) as implemented in Quantum Espresso package [4], which has been shown to accurately describe the phonon frequency under laser illumination [3]. The methodology for calculating the time-resolved atomic displacements amplitude ($Q$) of the $A_{1g}$ mode using Time-dependent density functional theory (TDDFT) combined with Ehrenfest molecular dynamics simulation has been previously described in detail in the previous article [5].

Limited by computing capability, we can only calculate the electron band structure of $Cs_2Ag_{0.5}Na_{0.5}InCl_6$ [Fig. S1(a)] which is similar to $Cs_2Ag_{0.4}Na_{0.6}InCl_6$. The electron band structure shows that the valence band is mainly composed of the 4*d* orbital electrons of Ag atoms and 3*p* orbital electrons of Cl atoms, while the 3*p* orbital electrons of Cl atoms and 5*s* orbital electrons of In atoms constitute the conduction band. The reorganization of excited-state structure induced by lattice distortion manifests in the variation of PES. Figure S1(b) shows that the calculated PESs of the excited state and ground state change with the configuration coordinates, which introduces the self-trapping energy $E_{st} = 0.09$ eV and lattice-deformation energy $E_d = 0.11$ eV. Herein, the corresponding average emission energy $E_{PL} = 2.18$ eV (570 nm).

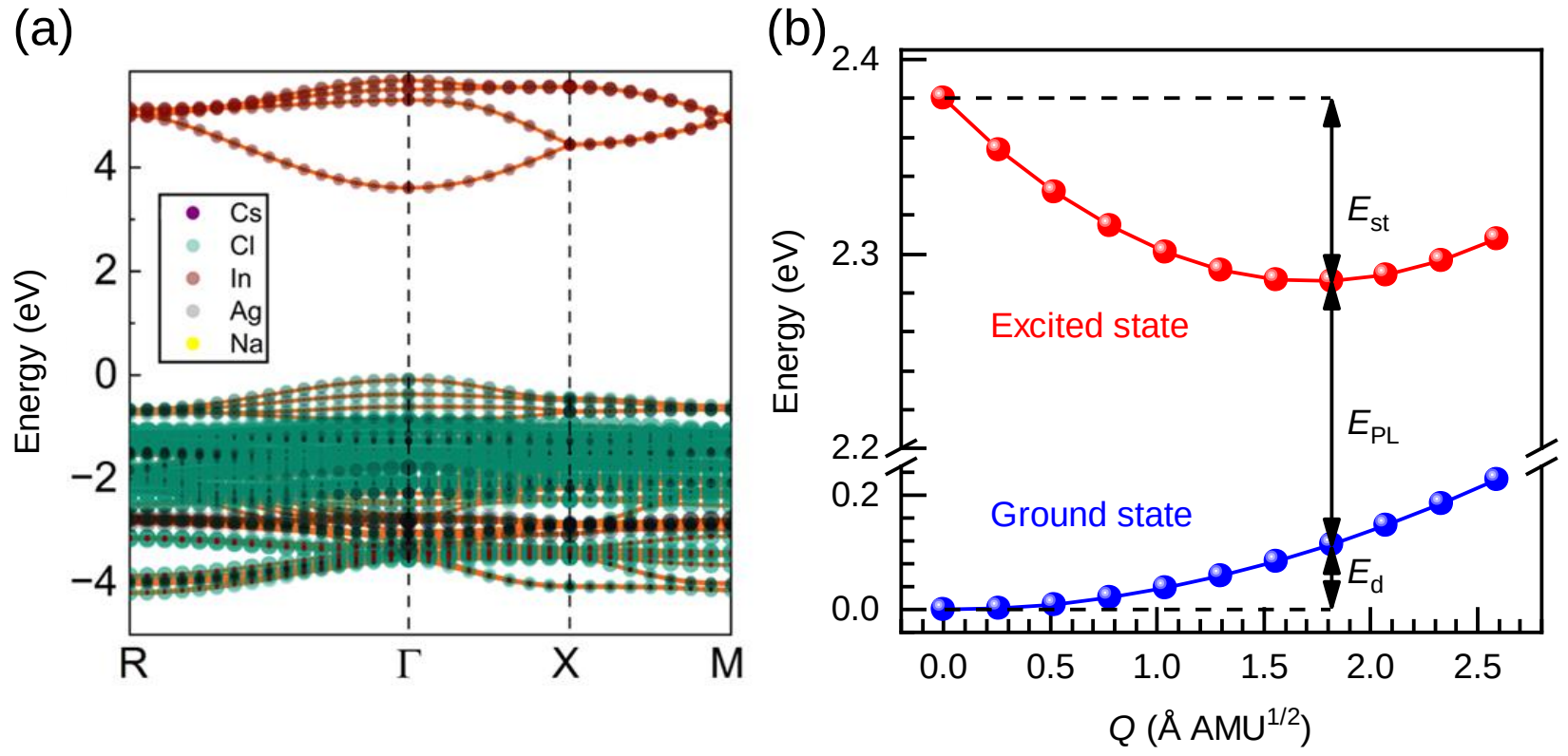

FIG. S1. Computational electron band structure and potential energy surface (PES) of $Cs_2Ag_{0.5}Na_{0.5}InCl_6$. (a) Electron band structure of $Cs_2Ag_{0.5}Na_{0.5}InCl_6$ calculated using PBE0 functional. The diameter of the scatter points shows the orbital-resolved partial density of states (PDOS) of the k-points along the high-symmetry path. (b) The schematic diagram of PES. $E_{st}$, $E_{PL}$, and $E_d$ correspond to the self-trapping energy, emission energy, and lattice-deformation energy, respectively.

## 2. The phonon dispersion and Raman tensors of $Cs_2Ag_{0.5}Na_{0.5}InCl_6$

Figure S2 shows the calculated phonon dispersion. The Raman-active $A_{1g}$ and $E_g$ modes are indicated using the blue solid lines. We are incapable of adequately assigning the Raman modes with relatively lower intensity around the three main peaks. Further study is needed to identify the remaining observed peaks.

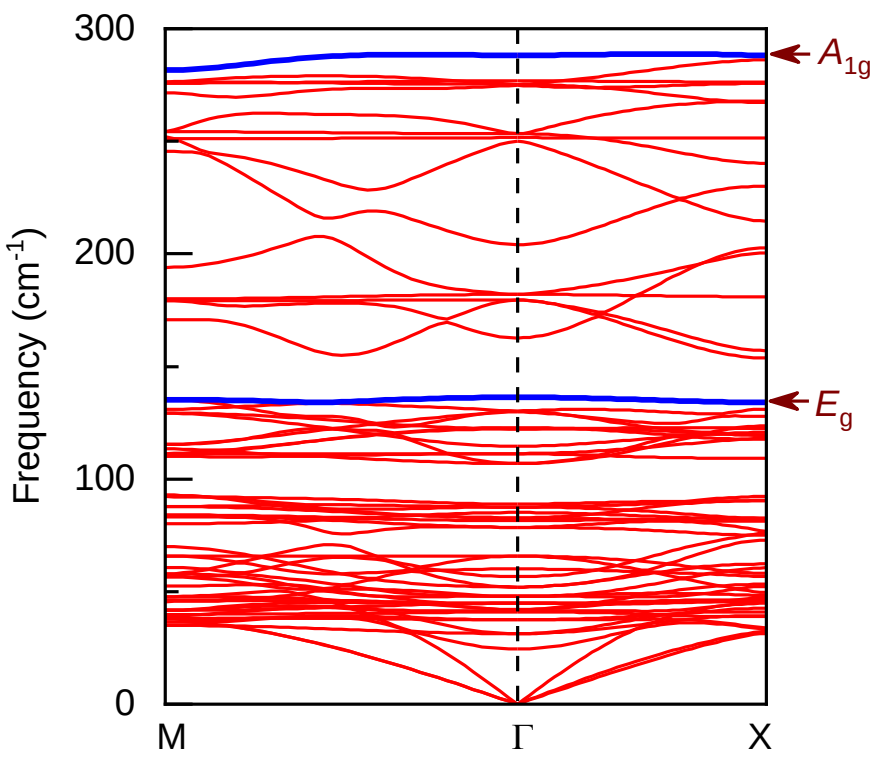

FIG. S2. Calculated phonon dispersion of $Cs_2Ag_{0.5}Na_{0.5}InCl_6$.

For the $O_h$ point group, there are three kinds of Raman-active modes. The corresponding Raman tensors are shown as follows [6]:

$$A_{1g} = \begin{pmatrix} a & 0 & 0 \\ 0 & a & 0 \\ 0 & 0 & a \end{pmatrix}, E_g = \begin{pmatrix} b & 0 & 0 \\ 0 & b & 0 \\ 0 & 0 & -2b \end{pmatrix}, E_g = \begin{pmatrix} -\sqrt{3}b & 0 & 0 \\ 0 & -\sqrt{3}b & 0 \\ 0 & 0 & 0 \end{pmatrix},$$

$$T_{2g} = \begin{pmatrix} 0 & 0 & 0 \\ 0 & 0 & d \\ 0 & d & 0 \end{pmatrix}, T_{2g} = \begin{pmatrix} 0 & 0 & d \\ 0 & 0 & 0 \\ d & 0 & 0 \end{pmatrix}, T_{2g} = \begin{pmatrix} 0 & d & 0 \\ d & 0 & 0 \\ 0 & 0 & 0 \end{pmatrix}. \quad \text{(S1)}$$

## 3. The laser lines and spectral resolutions under excitations with different energies

The Raman spectra were measured with Jobin-Yvon HR800 and T64000 systems equipped with a liquid nitrogen-cooled charge-coupled device (CCD) detector. The 100× objective lens (numerical aperture NA = 0.90) and 39× objective lens (NA = 0.49) were used to collect signals. 2400 gr/mm and 1800 gr/mm gratings were used in the Raman spectroscopy measurement. The PL spectra were collected by the Jobin-Yvon HR800 with 100 gr/mm grating at room temperature.

Figure S3 shows the full Raman spectra, FWHM of the laser lines, and the spectral resolution of the spectrometer. The laser lines were fitted using the Gauss function.

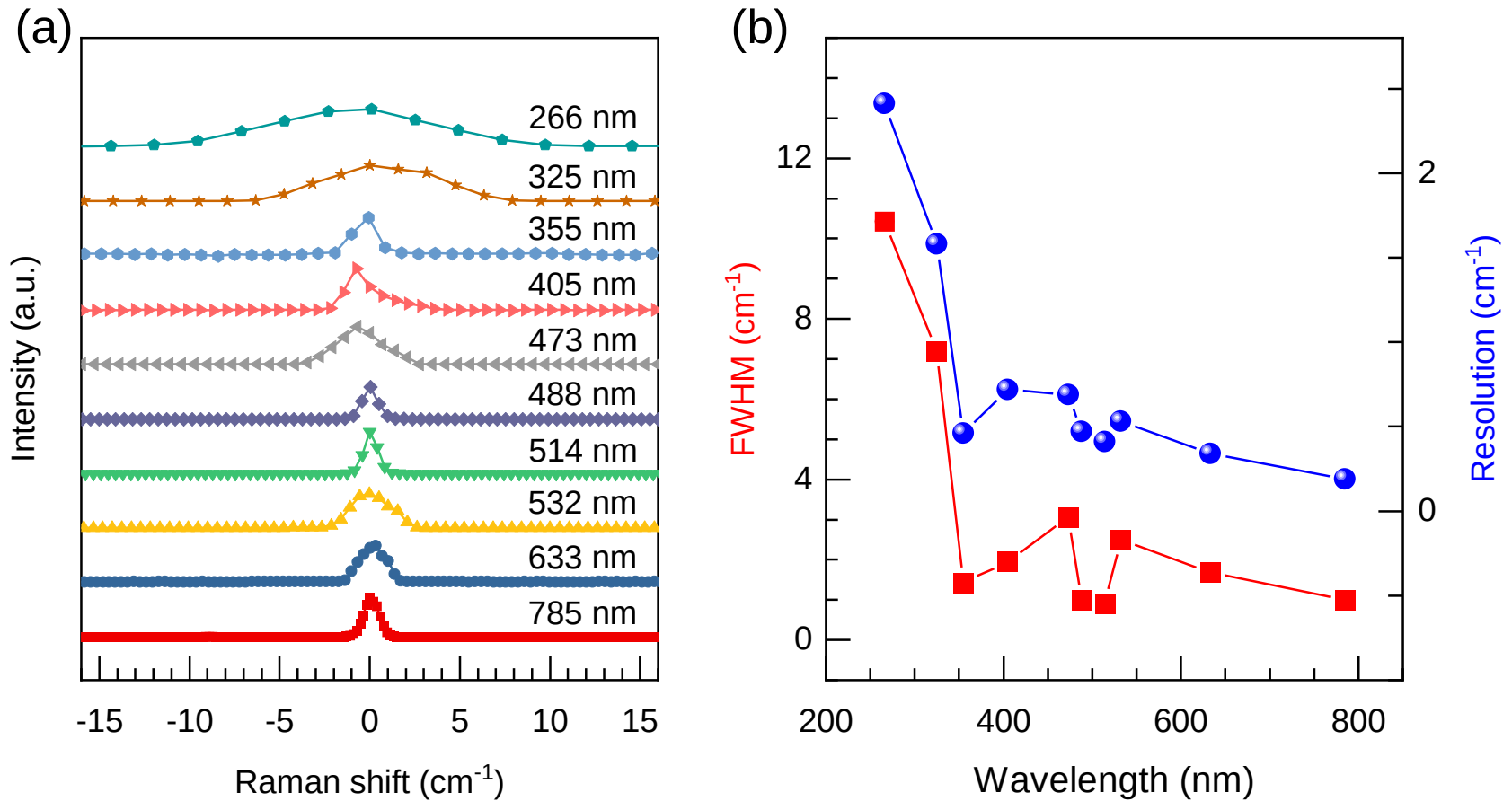

FIG. S3. The laser lines under different excitation energies. (a) The laser lines fitted using the Gauss function. (b) FWHM, and spectral resolution of the laser lines extracted from (a). The error bars correspond to the fitting error.

## 4. The contour plot of photoluminescence excitation (PLE) spectra of $Cs_2Ag_{0.4}Na_{0.6}InCl_6$

A contour plot of PLE spectra is shown in Fig. S4. The center wavelength of PL spectra in Fig. 3(c) was obtained by [7]

$$\overline{\lambda_f} = \frac{\int \lambda N(\lambda) d\lambda}{\int N(\lambda) d\lambda}, \tag{S2}$$

where $N(\lambda)$ is the intensity of the emission spectra at a certain wavelength $\lambda$. The FWHM of PL spectra was obtained by taking half the intensity of PL at the center wavelength.

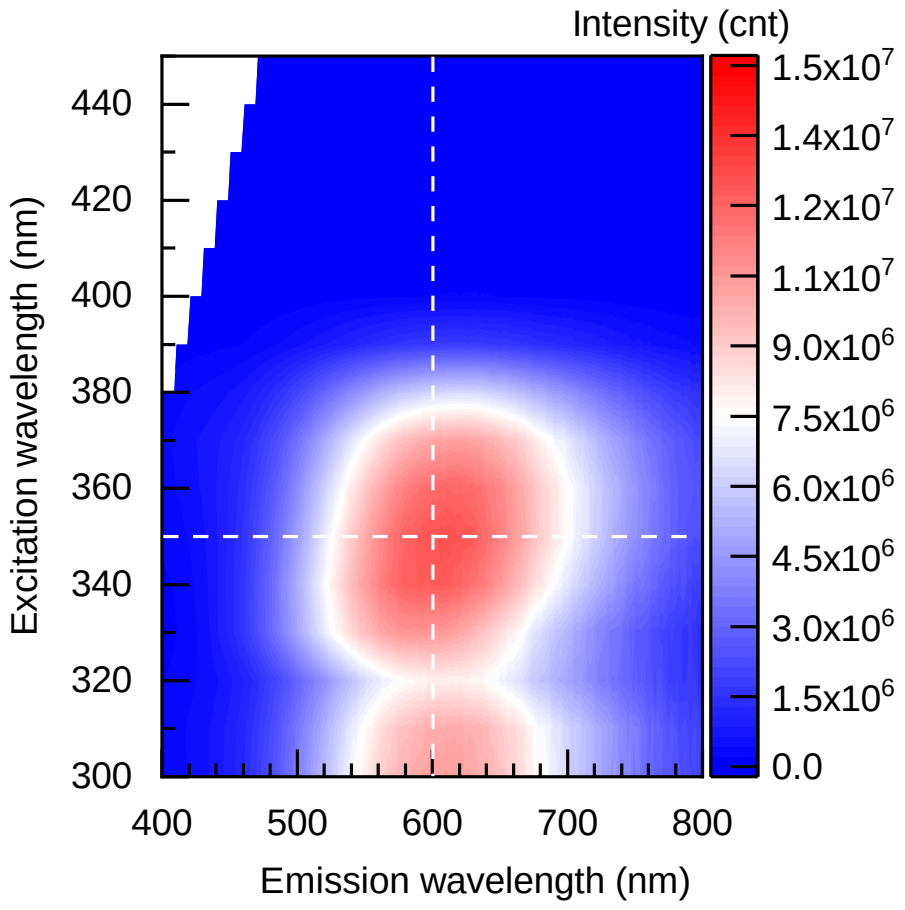

FIG. S4. The contour plot of PLE spectra of $Cs_2Ag_{0.4}Na_{0.6}InCl_6$.

## 5. The calculated PES and *S* factor of $Cs_2Ag_{0.4}Na_{0.6}InCl_6$

We theoretically calculated the excited state potential to evaluate the electron-phonon interaction (EPI) strength with increasing carrier concentration [Fig. S5(a)]. Based on the PESs of different photoexcited states, the Huang-Rhys factor $S_{fi,q}$ was obtained by the following formula [8]: $E_{fi} = W_f(Q_i) - W_i(Q_i) - \sum_q S_{fiq} \hbar\omega_q$, where $E_{fi}$ (emission energy) and $W_f(Q_i) - W_i(Q_i)$ (absorption energy) were extracted from Fig. S5(a), and $\hbar\omega_q$ (phonon energy of the $A_{1g}$ mode) was obtained from Raman spectra under 355 nm excitation.

As shown in Fig. S5(b), the $S$ factor as a function of carrier concentration can be fitted using $S = 0.06\,n^2$ (blue solid line).

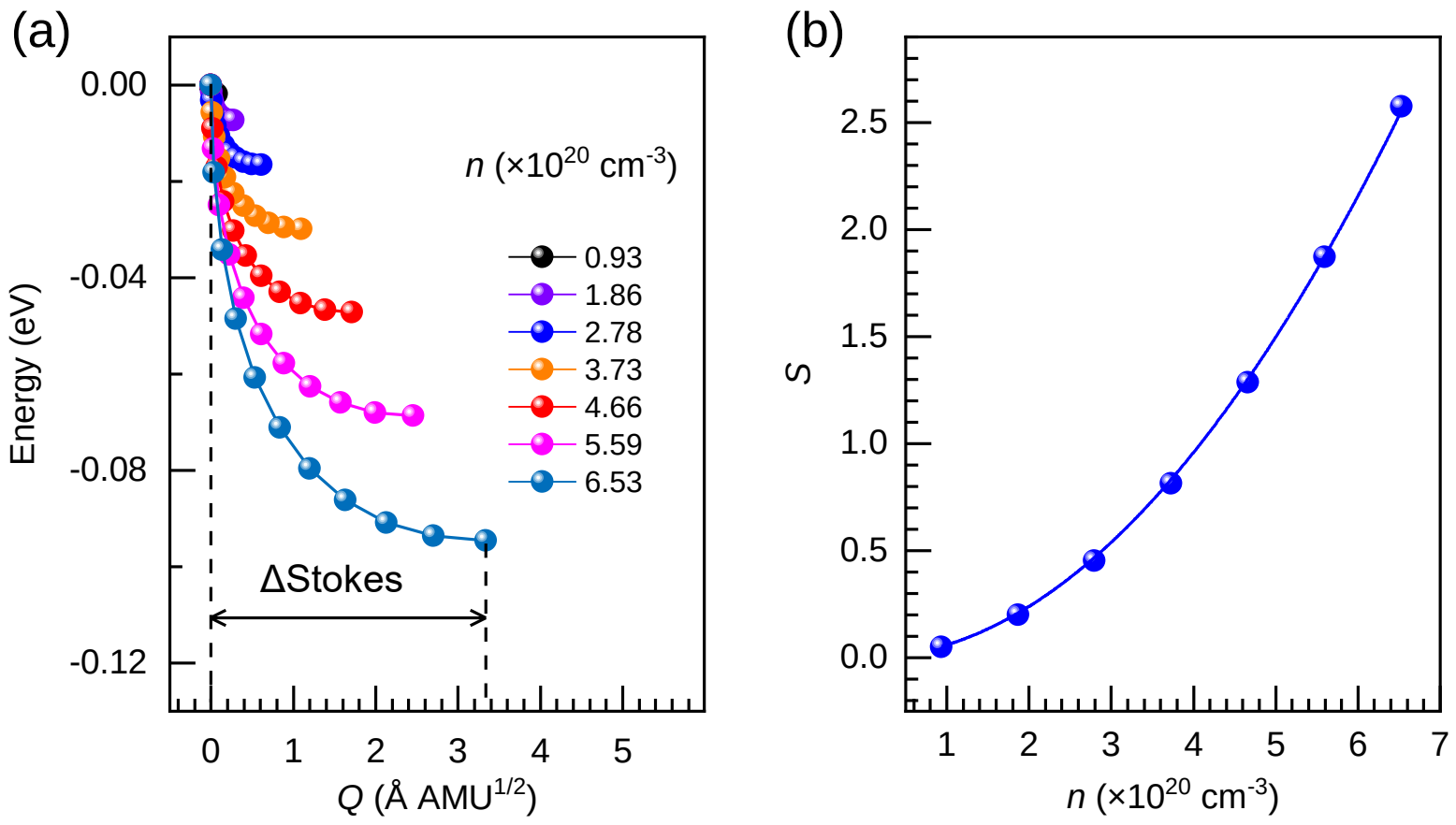

FIG. S5. The calculated PES and $S$ factor of $Cs_2Ag_{0.4}Na_{0.6}InCl_6$. (a) The calculated PES of the excited state of STE with increasing carrier concentration. (b) The calculated Huang-Rhys factor ($S$) with increasing carrier concentration. The blue solid line indicates the fitting result: $S = 0.06n^2$.

## 6. The Raman spectra under 532 nm, 325 nm, and 355 nm excitations

The Raman shifts and FWHMs of the $A_{1g}$ mode under the 532 nm (325 nm) excitation extracted from Fig. S6(a) [S6(b)] are shown in Fig. S6(d) [S6(e)]. Under the 532 nm excitation (without forming STE), the Raman shifts and FWHMs of the $A_{1g}$ mode almost remain unchanged with increasing power, while under the 325 nm excitation, they show similar behavior with the case under 355 nm excitation [Fig. 4(c)].

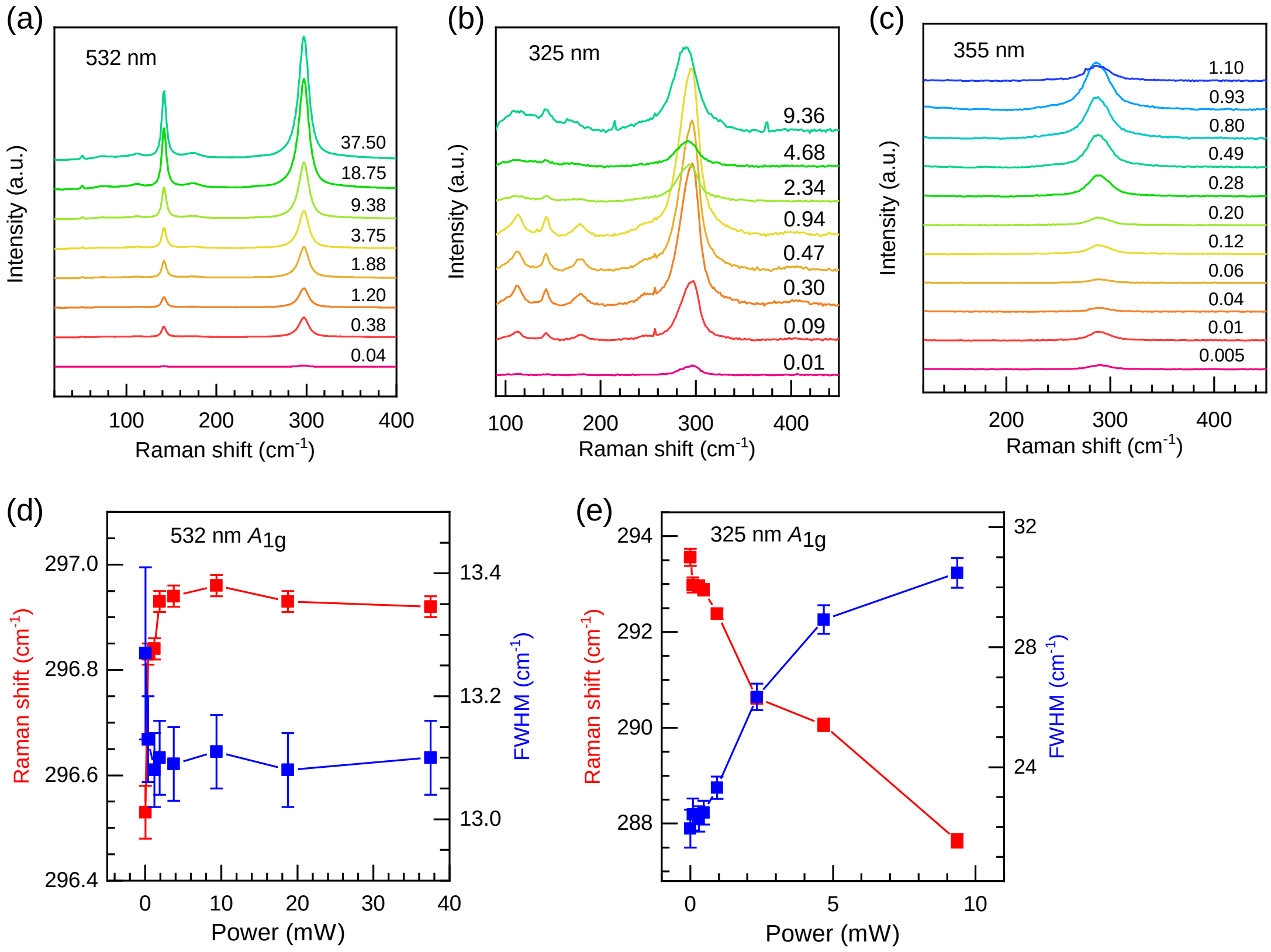

Fig. S6. The power-dependent Raman spectra of $Cs_2Ag_{0.4}Na_{0.6}InCl_6$. The power-dependent Raman spectra of $Cs_2Ag_{0.4}Na_{0.6}InCl_6$ under (a) 532 nm, (b) 325 nm, and (c) 355 nm excitation. The values in (a-c) indicate the power of excitation light, with the unit in mW. The Raman shifts and FWHMs of the $A_{1g}$ mode extracted from (a) and (b) under (d) 532 nm and (e) 325 nm excitation. The error bars correspond to the fitting error.